\documentstyle[preprint,aps,eqsecnum,epsfig]{revtex}
\def\mycomm#1{\hfill\break\strut\kern-3em{\tt ====> #1}\hfill\break}
\def\mycommNL#1{\strut\kern-3em{\tt ====> #1}\hfill\break}
\def\ds{\displaystyle}

\def\verydeepstrut{\vrule height 1.5ex depth 4.5ex width 0pt}

\def\thetap{\hbox{$\Theta^+$}}

\newcommand{\eqref}[1]{(\ref{#1})}   

\makeatletter
\def\hlinewd#1{\noalign{\ifnum0=`}\fi
\hrule \@height #1 \futurelet \reserved@a\@xhline}
\def\hwhiteline{\noalign
{\ifnum0=`}\fi\hrule
\@height 0pt\vskip 1.0ex\futurelet \reserved@a\@xhline}
\makeatother
\def\gray{\special{ps: 0.40 setgray}}
\def\black{\special{ps: 0.0 setgray}}

\newcommand{\mydraft}{
\newcount\timecount
\newcount\hours \newcount\minutes  \newcount\temp \newcount\pmhours

\hours = \time
\divide\hours by 60
\temp = \hours
\multiply\temp by 60
\minutes = \time
\advance\minutes by -\temp
\def\hour{\the\hours}
\def\minute{\ifnum\minutes<10 0\the\minutes
    \else\the\minutes\fi}
\def\clock{
\ifnum\hours=0 12:\minute\ AM
\else\ifnum\hours<12 \hour:\minute\ AM
\else\ifnum\hours=12 12:\minute\ PM
    \else\ifnum\hours>12
     \pmhours=\hours
     \advance\pmhours by -12
     \the\pmhours:\minute\ PM
     \fi
    \fi
\fi
\fi
}
\def\fullclock{\hour:\minute}
\begin{centering}
\gray
\font\Hugett  =cmtt12 scaled\magstep4
\hbox{\Hugett Draft:\today,\clock}
\black
\end{centering}
\vskip -1.7cm
$\phantom{a}$
} 

\def\beq#1{\begin{equation} \label{#1}}
\def\eeq{\end{equation}}

\def\ket#1{\left\vert #1\right\rangle}

\newskip\humongous \humongous=0pt plus 1000pt minus 1000pt

\newif\ifdtup

\begin{document}
{\tighten
\preprint
{}

\title{ A Diquark-Triquark Model for the
$KN$ Pentaquark}

\author{Marek Karliner\,$^{a,c}$\thanks{e-mail: \tt marek@proton.tau.ac.il}
\\
and
\\
Harry J. Lipkin\,$^{a,b}$\thanks{e-mail: \tt ftlipkin@weizmann.ac.il}
}
\address{ \vbox{\vskip 0.truecm}
$^a\;$School of Physics and Astronomy \\
Raymond and Beverly Sackler Faculty of Exact Sciences \\
Tel Aviv University, Tel Aviv, Israel\\
\vbox{\vskip 0.0truecm}
$^b\;$Department of Particle Physics \\
Weizmann Institute of Science, Rehovot 76100, Israel \\
High Energy Physics Division, Argonne National Laboratory \\
Argonne, IL 60439-4815, USA\\
\vbox{\vskip 0.0truecm}
$^c\;$Cavendish Laboratory, University of Cambridge, Madingley Road,
Cambridge, CB3~0HE, UK
}
\maketitle

\begin{abstract}%
 We propose a model for the recently
discovered $\Theta^+$ exotic $KN$ resonance as a novel kind of a
pentaquark with an unusual color structure: a $\overline{\hbox{\bf
3}}_c$ $ud$ diquark, coupled to $\hbox{\bf 3}_c$ $ud\bar s$
triquark in a relative $P$-wave. The state has
 $J^P=1/2^+$, $I=0$ and is
an antidecuplet of $SU(3)_f$.
A~rough mass estimate
of this pentaquark is close to experiment.
\end{abstract}%

\section{Introduction }

\subsection{Modeling the pentaquark: need both $qq$ and $q\bar q$ interactions}

The recent observation of the strange $\Theta^+$
 pentaquark\cite{Kyoto,Russia,Stepanyan:2003qr} with a mass of 1540 MeV
and a very small width $\sim$20 MeV
has generated a great deal of
interest. Although the original prediction of an exotic $KN$ resonance was
obtained within the framework of the Skyrme model \cite{SMKNa,SMKNb}, there is
an obvious and urgent need to understand what $\Theta^+$ is in the quark language
\cite{Stancu:2003if}.

An additional nontrivial challenge for the quark interpretation \cite{JaffePC}
is that whereas the Skyrme model predicts that $\Theta^+$ has positive parity,
the ``standard" pentaquark involves 5 quarks in an $S$-wave and therefore has
negative parity. As of now, there is no clearcut experimental information on
the $\Theta^+$ parity, but if it is positive, clearly one must have one unit of
orbital angular momentum and this makes the calculation difficult.

The most straightforward interpretation of the \thetap\ in terms
of quarks is that it is a $uudd\bar s$ pentaquark, so it has both
$qq$ and $q\bar q$ interactions. At present  it is not possible to
compute the properties of such a state from first principles, so
it is necessary to use a model which is known to reliably deal
with both types of interactions.

The quark model we use provides such a unified treatment of both
types of interactions in mesons and baryons. Pioneered by Sakharov
and Zeldovich\cite{SakhZel}, it has subsequently been extended and
motivated within the framework of QCD by De Rujula, Georgi and 
Glashow \cite {DGG}, in
terms of color-magnetic interaction model for the hyperfine
interaction, and augmented by Jaffe's color-spin
algebra\cite{Jaffe} for multiquark systems. 

To provide a basis for the credibility for our use of the model
and to prepare the tools for the analysis of the pentaquark,
we now briefly review 
and update the successes of the model for a unified treatment of mesons and
baryons of all flavors. 

\subsection {Summary of successful mass relations from hadrons
containing no more than one strange or heavy quark.}

Early evidence that mesons  and baryons are made of the same quarks was
provided by the remarkable successes of the constituent
quark model \cite{SakhZel}, in which static  properties and low 
lying excitations of both
mesons and baryons are described as simple composites of asymptotically free
quasiparticles  with a flavor dependent linear mass term and hyperfine
interaction, yielding a unified mass formula for both mesons and baryons
\beq {sakzel}
M = \sum_i m_i + \sum_{i>j}  {{\vec{\sigma}_i\cdot\vec{\sigma}_j}\over{m_i\cdot
m_j}}\cdot v^{hyp}_{ij}
\end{equation}
where $m_i$ is the
effective mass of quark $i$, $\vec{\sigma}_i$ is a quark  spin operator and
$v^{hyp}_{ij}$ is a hyperfine interaction with different strengths but the same
flavor dependence and we have added the explicit flavor dependence of the
hyperfine interaction \cite {DGG}.

The effective quark mass appears in two different terms 
in eq.~\eqref{sakzel}\,: as an additive term and
in the denominator of the hyperfine interaction.
In all the relations for
masses and magnetic moments obtained in the light ($uds$)  flavor sector, and
for hadrons containing no more than one heavy or strange quark,
agreement with experiment has been obtained by assuming that the values of the
effective quark masses in these two terms has been the same and that the values
are the same for mesons and baryons. Both the mass difference and the mass
ratio between two quarks of different  flavors were found  to have the same
values to a good approximation when they are bound to a nonstrange antiquark to
make a meson and  bound to a nonstrange diquark to make a baryon.

For
example, the effective quark mass difference $m_s-m_u$ is found to
have the same value $\pm 3\%$ and the mass ratio $m_s/m_u$ the
same value $\pm 2.5\%$, when calculated  from baryon masses and
from meson masses\cite{SakhZel,ICHJLmass,HJLMASS}, with a simple
recipe for removing the hyperfine contribution. Thus the mass difference 
of two quarks, denoted by $Q$ and $q$, can be obtained 
from meson masses,
\beq{szdif}
\langle m_Q - m_q\rangle_{Mes} = 
{3 M_{{\cal V}_{Q \bar u}} + M_{{\cal P}_{Q \bar u}} \over 4}
-
{3 M_{{\cal V}_{q \bar u}} + M_{{\cal P}_{q \bar u}} \over 4}
\end{equation}
where ${\cal V}_{Q \bar u}$ and ${\cal P}_{Q \bar u}$ denote 
the vector and pseudoscalar mesons with the constituents $Q \bar u$, etc.
The same observable can also be obtained from baryon masses,
\beq{szdifbar} \langle
m_Q - m_u\rangle_{Bar} = M_{\Lambda_Q} -M_N
\end{equation}
so that for $Q=s$ and $q=u$ one has
\beq{SZeq}
\begin{array}{ccrclcc}
\langle m_s{-}m_u \rangle_{Bar} &=&M_\Lambda &-&M_N&=& 177\,{\rm MeV}
\\
\mbox{}\\
\langle m_s{-}m_u \rangle_{Mes} &=&
\ds  {3M_{K^*}  +M_K\over 4} &-&
\ds {3M_\rho -M_\pi \over 4 } 
&=&
179\,{\rm MeV}
\end{array}
\end{equation}

 The same approach has been applied to heavy flavors
\cite{PBIGSKY,NewPenta} with excellent results.
In this context we note a new relation \cite{NewPenta},
showing the common nature of the
hyperfine interactions in mesons and baryons of different flavors,
\beq{SigLam4}
{{M_{\Sigma_c} - M_{\Lambda_c}}\over{M_{\Sigma} - M_\Lambda}}=2.16
\,\,\approx\,\,
{{(M_\rho - M_\pi)-(M_{D^*}-M_D)}
\over
{(M_\rho - M_\pi)-(M_{K^*}-M_K)}}
=2.10
\end{equation}

We exhibit this success in more detail, by showing that
mass differences and mass ratios are fit
with a single set of quark masses, chosen to give an eyeball fit 
to the baryon mass differences and to fit the isoscalar nonstrange 
baryon magnetic moment
\beq{isomag}
\mu_p+\mu_n=
2M_{\scriptstyle p}\cdot {{Q_I}\over{M_I}}
={2M_N\over M_N+M_\Delta}=0.865 \,{\rm n.m.}
\qquad(\hbox{EXP}  =
0.88 \,{\rm n.m.})
\verydeepstrut
\end{equation}
where
$ Q_I= \ds {1\over 2}\cdot \left( {2\over3} - {1\over 3} \right) =
 {1\over 6} $ \ and \
$M_I= \ds {1\over 6}\cdot \left( M_N + M_\Delta \right)$
denote the charge and mass, respectively, of an effective
``isoscalar nonstrange quark".\footnote{Note the implicit assumption
that in $M_I$ the contribution
of the hyperfine interaction is cancelled between the nucleon
and the $\Delta$.}
The quark masses chosen for the fit were
\beq{qmass}
m_u =  360
\hbox{\ MeV};\qquad
m_s=  540
\hbox{\ MeV};\qquad
m_c= 1710
\hbox{\ MeV};\qquad
m_b=  5050
\hbox{\ MeV}\,.
\end{equation}

The results are shown in Table II below.

\vskip1cm
\vbox{
{\centerline{\bf{Theoretical and Experimental Hadron Mass Differences and Ratios}}}

\vskip 1.truecm

\centerline{\bf{TABLE II-A - Hadron Mass Differences}}
$$ \vcenter{
\halign{${#}$\quad
        &${#}$\quad
        &${#}$\quad
        &${#}$\cr
 {\rm \ \ \ Mass~Difference} &{\rm Theoretical}&{\rm Experimental}&{\rm Experimental} \cr
    & {\rm From~eq.~(\ref{qmass})}   & {\rm From~Mesons}~(X{=}d) & {\rm From  ~ Baryons}~(X{=}ud)\cr
m_s{-}m_u = M(sX) {-} M(uX)\ \  &  \,\,180  & \,\, 179   &   \,\,  177     \cr
m_c{-}m_u = M(cX) {-} M(uX) & 1350  & 1360  &   1346    \cr
m_b{-}m_u = M(bX) {-} M(uX) & 4690 &  4701   &    4685    \cr
m_c{-}m_s = M(cX) {-} M(sX) & 1170 &  1180   &  1169   \cr
m_b{-}m_s = M(bX) {-} M(sX) & 4510 &  4521  &     4508   \cr
m_b{-}m_c = M(bX) {-} M(cX) & 3340  & 3341  &   3339     \cr
}}   $$
}
\par

\vskip 1.0truecm

\centerline{\bf{TABLE II-B - Quark Mass Ratios}}
\nopagebreak
$$ \vcenter{
\halign{${#}$\quad
        &${#}$\quad
        &${#}$\quad
        &${#}$\cr
 {\rm Mass~Ratio} & {\rm Theoretical} & {\rm Experimental}& {\rm Experimental} \cr
   & {\rm From~eq.~(\ref{qmass})}  & {\rm From~Mesons}~(X=d)&{\rm From~Baryons}~ (X=ud)\cr
\ m_s/m_u &  1.5   &   1.61 &   1.53      \cr
\ m_c/m_u &  4.75   &  4.46  &  4.36 \cr
\ m_b/m_u &  14.0  & 13.7  &   ?    \cr
\ m_c/m_s &  3.17  & 2.82  &  2.82  \cr
\ m_b/m_s &  9.35 &  8.65  &  ?\cr
\ m_b/m_c &  2.95  & 3.07  &  ?  \cr
}}   $$
\hfill\break
\par

While we await for QCD calculations  to explain these striking experimental
facts from first principles,  we use the method to analyse the pentaquark
color structure and to estimate its mass.

\section{The dynamics of a diquark-triquark pentaquark} 
Most quark model treatments of multiquark spectroscopy use the color-magnetic 
short-range hyperfine interaction\cite{DGG} as the dominant mechanism for
possible binding. The treatment of exotic color configurations not found in
normal hadrons is considerably simplified by the use of
color-spin $SU(6)$ algebra \cite{Jaffe}. The
the  hyperfine interaction between 
two quarks denoted by $i$ and $j$ is then written as 
\beq{vhypI}
V_{hyp} =-V(\vec \lambda_i \cdot \vec \lambda_j)(\vec \sigma_i \cdot \vec \sigma_j)
\end{equation} 
where $\vec \lambda$ and $\vec \sigma$ denote the generators of $SU(3)_c$ and
the Pauli spin operators, respectively. 
The sign and magnitude of the strength of the hyperfine interaction are 
normalized by $\Delta$-$N$ mass splitting.
The quark-quark interaction \eqref{vhypI} 
is seen to be attractive
in states symmetric in color and spin where 
$(\vec \lambda_i \cdot \vec \lambda_j)$ and $(\vec \sigma_i \cdot \vec
\sigma_j)$ have the same sign 
and repulsive in antisymmetric states whee they have opposite signs.
This then leads to the "flavor-antisymmetry" principle\cite{Lipflasy}: 
the
Pauli principle forces two identical fermions at short distances to be in a
state that is antisymmetric in spin and color where the hyperfine interaction
is repulsive.  Thus {\em
the hyperfine interaction is always repulsive between two
quarks of the same flavor, such as the like-flavor $uu$ and $dd$  pairs in the
nucleon or pentaquark.}

This flavor antisymmetry suggests that the bag or single-cluster models
commonly used to treat normal hadrons may not be adequate for multiquark
systems. In such a state, with identical pair correlations for all pairs in the
system, all same-flavor quark pairs are necessarily in a higher-energy
configuration, due to the repulsive nature of their hyperfine interaction. 
The
$uudd\bar s$ pentaquark is really a complicated five-body system  where the
optimum wave function to give minimum color-magnetic energy can require
flavor-dependent spatial pair correlations for different pairs in the system;
e.g. that keep the like-flavor $uu$ and $dd$ pairs apart,  while minimizing the
distance and optimizing the color couplings within the  other pairs. 

We consider here a possible model for a strange pentaquark that implements 
these ideas
by dividing the system into two color non-singlet clusters
which separate the pairs of identical flavor.
The two clusters, a $ud$ diquark and a $ud\bar s$ triquark, are in a 
a relative $P$-wave and
are separated by a distance larger than the range of the color-magnetic force
and are kept together by the color electric force. Therefore the color hyperfine
interaction operates only within each cluster, but is not felt between the
clusters, as shown schematically in Fig.~1.

\hfill\break
\hfill\break
\centerline{\epsfig{file=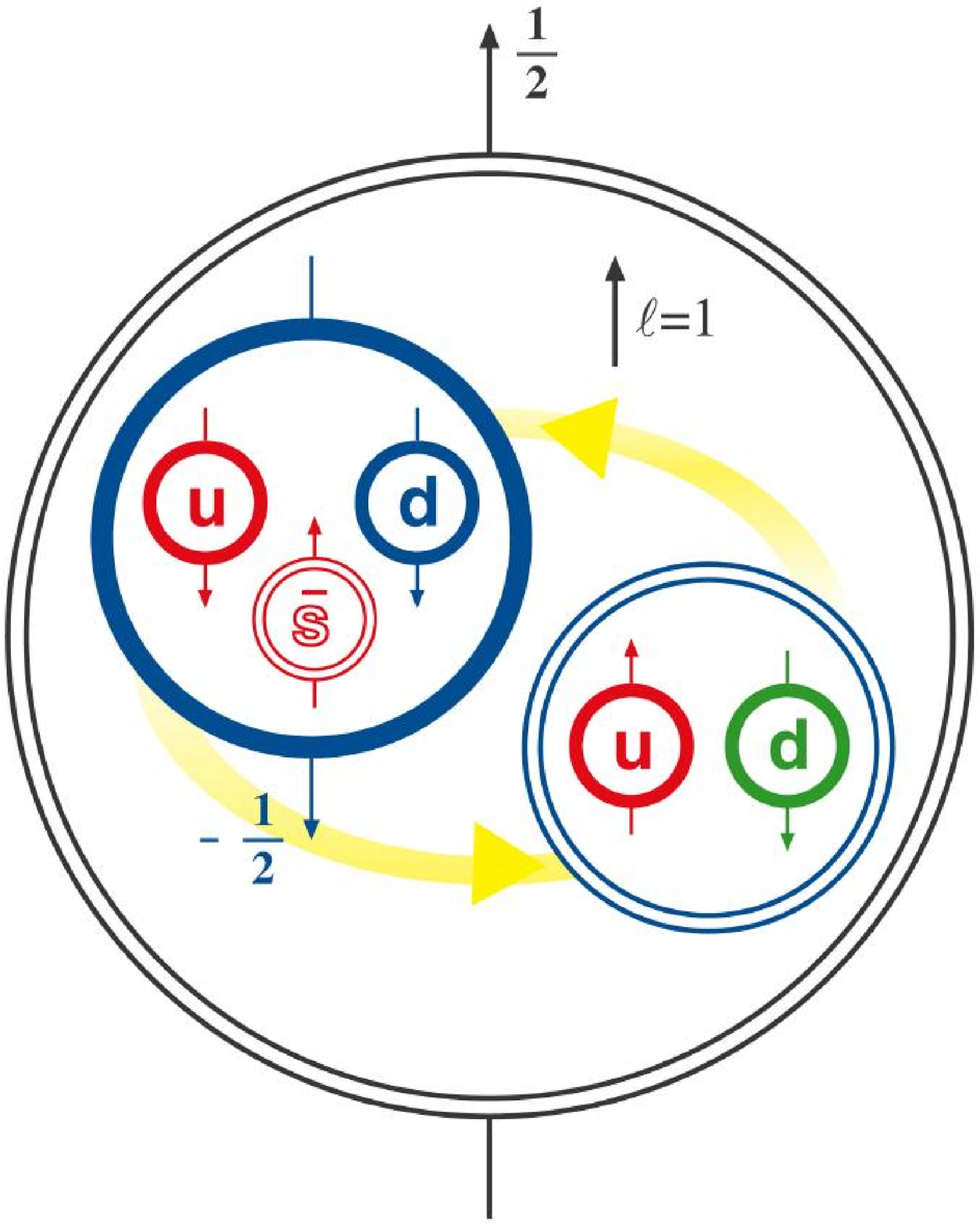,width=10.0cm,angle=0}}
\centerline{\it \small Fig. 1. The diquark-triquark configuration of the
$uudd\bar s$ pentaquark.}
\hfill\break

The $ud$ diquark is in the $\bar {\hbox{\bf 3}}$
of the color $SU(3)$  and in  the $\bar {\hbox{\bf 3}}$ of the flavor $SU(3)$
and has $I=0, S=0$, like the $ud$
diquark in the $\Lambda$. It is in the symmetric {\bf 21} of the color-spin $SU(6)$
and is antisymmetric in both spin and color.

The {\bf 21} representation of $SU(6)$ contains a color antitriplet with spin 0
and a color sextet with spin 1.

The $ud$ in the $ud\bar s$ triquark is in {\bf 6} of $SU(3)_c$,
in $\bar{\hbox{\bf 3}}$ of $SU(3)_f$ and has $I=0, S=1 $.
It is also in the symmetric
{\bf 21} of the color-spin $SU(6)$, but is symmetric in both spin and color.

The triquark consists of the diquark and antiquark coupled to an $SU(3)_c$
triplet and has $I=0, S=1/2$. It is in the fundamental {\bf 6} representation of
the color-spin $SU(6)$. It is in a $\bar{\hbox{\bf 6}}$ of  $SU(3)_f$.

We now define the classification of the diquarks with spin $S$,
denoted by $\ket{(2q)^S}$ and the triquark, denoted by
 $\ket{(2q\bar s)^{1\over2}}$,
in a conventional notation $\ket{D_6,D_3,S,N}$
\cite{Patera,Sorba}
where $D_6$ and $D_3$ denote the dimensions of the color-spin $SU(6)$ and
color $SU(3)$ representations in which the multiquark states are classified,
$S$ and $N$ denote the total spin and the number of quarks in the system,
\def\smallhalf{\hbox{${1\over2}$}}
\begin{eqnarray}
\ket{\,\,(2q)^1\,}          &=& \ket{21,6,1,2}    \cr
\ket{\,\,(2q)^0\,}          &=& \ket{21,\bar 3,0,2}   \\
\ket{(2q\bar s)^{1\over2}}    &=& \ket{\,\,6,3,\smallhalf,3}    \nonumber
\end{eqnarray}

A standard treatment using the $SU(6)$ color-spin algebra\cite{Patera,Sorba}
gives the result in the SU(3)-flavor symmetry limit that  the hyperfine
interaction is stronger by ${1\over6}(M_\Delta-M_N)$ for the diquark-triquark
system than for the kaon-nucleon system,
\begin{equation} [V(2q\bar s^{1\over2}) + V(2q^0)] - [V(K )+ V(N)]=
-{1\over6}(M_\Delta-M_N)  \approx {-}50 {\rm MeV} \label{MassShift}
\end{equation} The physics here is simple. The spin-zero diquark is the same as
the diquark in a $\Lambda$ and has the same hyperfine energy as a nucleon. A
triquark with one  quark coupled with the $\bar s$ antiquark to spin zero has
the same hyperfine energy as a kaon but no interaction with the other quark.
The triquark coupling used here allows the $\bar s$ antiquark to interact with
both the $u$ and $d$ quarks and gain hyperfine energy with respect to the case
of the kaon. For an isolated triquark such a configuration is of course
forbidden, since it a color nonsinglet, but here it is OK, since the triquark
color charge is neutralized by the diquark.

We see that had it not been for the cost of the $P$-wave excitation,
the triquark-diquark system would be somewhat more 
bound than a kaon and a nucleon. 
The diquark and triquark will have a color electric interaction
between them which is identical to the quark-antiquark interaction in a meson.
If we neglect the finite sizes of the diquark and triquark we can compare
this system with analogous mesons. We can use the effective quark masses
(\ref{qmass}) that
fit the low-lying mass spectrum\cite{SakhZel,NewPenta}
to find a very rough estimate
\beq{qmass2}
m_{diq} =  720
\hbox{\ MeV};\quad
m_{triq}=  1260
\hbox{\ MeV};\quad
m_r(di\hbox{-}tri)= 458
\hbox{\ MeV}
\,.
\end{equation}
where $m_{diq}$ and
$m_{triq}$ denote the
effective masses of the diquark and triquark, $m_r(di\hbox{-}tri)$ denotes 
the reduced
mass for the relative motion of the diquark-triquark system.

A crucial observation is that the diquark-triquark system may not exist in
a relative \hbox{$S$-wave}. 
This is because in $S$-wave the hyperfine interaction acts not
only within the clusters but also between them. The repulsive terms may then win and
the would be $S$-wave gets rearranged into the usual $K N$ system. The situation
is different in a $P$-wave, because then the diquark and the triquark are separated
by an angular momentum barrier and the color-magnetic interactions operate only
within the two clusters.  The price is the $P$-wave excitation energy.

We can obtain a rough estimate of this $P$-wave excitation energy, using
experimental information about the excited states of $D_s$, since 
the reduced mass of the $c\bar s$ system 
used to describe the
internal structure of the $D_s$ spectrum is 410 MeV,
quite close to that of the diquark-triquark system.

It has been proposed that the recently discovered extremely narrow
resonance $D_s$(2317) \cite{Aubert:2003fg,CLEO,BELLE-Ds}
is a $P$-wave excitation \cite{Bardeen:2003kt}
of the ground state $0^-$ $D_s$(1969). 
If so, the \,350 MeV excitation energy then
consists of a $P$-wave contribution, on top of a contribution from 
color hyperfine splitting. We can estimate the net $P$-wave
excitation energy \,$\delta E^{P-wave}$\,
by subtracting the $c$-$s$ hyperfine splitting
obtained from the mass difference between $D_s^*$ and $D_s$,
\begin{equation}
\delta E^{P-wave} \approx  350 -
(m_{D_s^*} - m_{D_s}) = 207\ \rm MeV
\end{equation}

From eq.~(\ref{MassShift}) we infer that without the $P$-wave
excitation energy the diquark-triquark mass is
$m_N+m_K-{1\over6}(M_\Delta-M_N)\approx 1385$ MeV, so that the
total mass of the $P$-wave excitation of the diquark-triquark
system is expected to be \beq{PentaMass} M_{di\hbox{-}tri} \approx
1385+207 = 1592 \ \rm MeV\,, \eeq about 3\% deviation from the
observed mass of the $\Theta^+$ particle. It should be kept in
mind, however, that this is only a very rough qualitative estimate
and this close agreement might well be fortuitous, as there are
several additional model-dependent effects which should be taken
into account: the reduced mass of $D_s$ is $\sim 12\%$ lower than
$m_r(di\hbox{-}tri)$, we don't know the spatial wave functions and
we have neglected the spatial extent of the diquark and triquark
and possible molecular Van-der-Waals interactions spatially
polarizing the two, breaking of flavor $SU(3)$, etc.

In addition to the parity and the mass, we also note that our model
naturally gives a state with isospin zero
because both the diquark and triquark have $I=0$. The isospin has not yet been
determined experimentally, but no isospin partners of the $\Theta^+$ have been
found and the Skyrme also predicted $I=0$.
This should be contrasted
with attempts to envision the $\Theta^+$ as a $KN$ molecule
in a $P$-wave \cite{Mitya}, which  have a problem in getting rid of the $I=1$ state.

Our model also naturally predicts that the $\Theta^+$ is in
an antidecuplet of $SU(3)$ flavor. The diquark is a $\bar{\hbox{\bf 3}}$, the
triquark a $\bar{\hbox{\bf 6}}$ and
in $SU(3)$
$\bar{\hbox{\bf 3}} \otimes \bar{\hbox{\bf 6}}=
\overline{\hbox{\bf 10}} \oplus  {\hbox{\bf 8}}$
and only $\overline{\hbox{\bf 10}}$ has the right strangeness.
$KN$ is $\hbox{\bf 8} \otimes \hbox{\bf 8}$
in $SU(3)_f$ and contains {\bf 27} with an isovector with the
right strangeness, in addition to an antidecuplet.
The antidecuplet prediction is again
in agreement with the Skyrme model.

Since $M_{di\hbox{-}tri}$ is above the $KN$ threshold,
the system will eventually decay to
$KN$, but the orbital angular momentum barrier and
the required color rearrangement will make such a decay relatively slow,
possibly explaining the observed narrow width of the $\Theta^+$.

\section{Effects of flavor symmetry breaking}

   The treatment above assumes flavor symmetry; i.e. that all quarks and the
antiquark have the same mass. We examine the symmetry breaking for a
pentaquark $\Theta(uudd\bar Q)$, with an antiquark of flavor $Q$, with a mass
different from the mass of the four quarks. This applies not only to the
$\Theta^+$ with a strange
antiquark but also to states  with heavier antiquarks. The mass difference
between the pentaquarks $\Theta(uudd\bar Q)$ and  $\Theta(uudd\bar q)$,
where the antiquark $\bar q$ has the same mass
as the $u$ and $d$, is just the sum of the differences in the masses and
in the hyperfine energies of the antiquarks,

The same treatment which leads to eq.~\eqref{MassShift} now gives for the 
total hyperfine interaction in our diquark-triquark model for 
$\Theta_Q$:
\begin{equation}
V(\Theta_Q)={-}(7+13\,\zeta)\cdot{m_\Delta-m_N \over 12}
\label{VKL}
\end{equation}
where $\zeta \equiv m_u/m_Q$.
This should be compared with the hyperfine energy of the nucleon and
the $u\bar Q$ meson,
\begin{equation}
V(N)+V(u\bar Q)= {-}(1+2\,\zeta)\cdot{m_\Delta-m_N \over 2}
\label{VNmeson}
\end{equation}
so that the difference in the hyperfine interaction between the
diquark-triquark configuration and the $N$ $u\bar Q$ system is 
\begin{equation}
V(\Theta_Q)- [V(N)+V(u\bar Q)] = 
{-}(1+\zeta)\cdot {m_\Delta-m_N \over 12}
\label{VKLshift}
\end{equation}
For $\zeta=1$ we recover the result in eq.~\eqref{MassShift}.
For a realistic $m_s$, we take $\zeta=2/3$, obtaining a small correction
\begin{equation}
V(\Theta^+)- [V(N)+V(K)] =
{-}{5\over 36}\cdot (m_\Delta-m_N)= {-}42\ \hbox{MeV}.
\label{VKLshiftThetaPlus}
\end{equation}

The same approach can be used to treat pentaquarks with $\bar c$ and $\bar b$ 
antiquarks\cite{HeavyPenta}.

We now examine the $\Xi^*(I=3/2)$ , which has the quark constituents
$(uuss\bar d)$.and the same mass as the $\Theta^+$ in the SU(3) limit.
For this case we set $\zeta = m_u/m_s=(2/3)$ . 

For the hyperfine 
 interaction in the $us$ diquark with spin 0 and $\zeta=(2/3)$  we obtain,
\beq{usdiq}
V(us) = {-} {\zeta \over 2}\cdot (M_\Delta-M_N) 
= {-}{1\over3}\cdot (M_\Delta-M_N) 
\end{equation}
For the $(us \bar d)$ triquark hyperfine 
 interaction we obtain
\beq{usbtriq}
 V(us\bar d)  ={-}(13+15\,\zeta)\cdot{m_\Delta-m_N \over 24} =
 -{23\over24}\cdot (M_\Delta-M_N) 
 \end{equation}
Here the quark-quark interaction is modified by a factor $\zeta$, while the
quark-antiquark interaction is modified by a factor $(1 + \zeta)/2$,  since
only half of the two quarks is strange. Putting \eqref{usdiq} and 
\eqref{usbtriq}, we obtain the total hyperfine interaction in 
$\Xi^*(I=3/2)$
\beq{xi}
 V(\Xi^*(I=3/2)) ={-}(13+27\,\zeta)\cdot{m_\Delta-m_N \over 24} = 
  -{31\over24}\cdot (M_\Delta-M_N)  
\end{equation}
The difference between the $\Xi^*(I=3/2)$ and
$\Theta^+$ hyperfine interactions is then 
\beq{klthetaxi}
\delta V_{hyperfine} \equiv V(\Xi^*(I=3/2))   - V(\Theta^+) 
  =(1 -\zeta)\cdot{m_\Delta-m_N \over 24} = 
 {M_\Delta-M_N\over 72}  = 4.2\ \hbox{MeV}\,.
\end{equation}

The $\Xi^*(I=3/2)$ mass is obtained from the experimentally known 
mass of $\Theta^+$ by adding the quark mass difference $(m_s - m_u)$  
and the hyperfine energy difference,
\beq{klthetaxi2}
 M_{\Xi^*(I=3/2)} = M_{\Theta^+} + (m_s - m_u) 
   +\delta V_{hyperfine}= 1540 + 178 + 4 =1722\ \hbox{MeV}.    
\end{equation}
Since $M_\Xi + M_\pi = 1460    \ \hbox{MeV} $,
the mass of the $\Xi^*(I=3/2)$ is about 260 MeV above threshold.

\section*{Summary and Conclusions}

We propose the interpretation  of the recently discovered $\Theta^+$ exotic
$KN$ resonance as a novel kind of a pentaquark, involving a recoupling of
the five quarks into a diquark-triquark system in non-standard color
representations. We estimate the $\Theta^+$ mass using
the simple generalized Sakharov-Zeldovich mass
formula which holds with a single set of effective quark mass values for
all ground state mesons and baryons having no more than one  strange or
heavy quark.

Our rough numerical estimate indicates that such a color recoupling
might put the pentaquark mass in the right ballpark of the experimentally observed
$\Theta^+$ mass. Our model naturally predicts that $\Theta^+$ has spin 1/2,
positive parity, is an isosinglet and is an antidecuplet in $SU(3)_f$.
We calculate the effect of $SU(3)_f$ symmetry breaking and the mass splitting
between the $\Theta^+$ and another member of the antidecuplet,
the $\Xi^*(I=3/2)$.

Regardless of the specific details of the model, we have addressed
the problem what kind of a five-quark configuration can describe
the $\Theta^+$. We have shown that our new diquark-triquark model
with color recoupling gives a lower mass than the simplest
$uudd\bar s$ and it looks promising. The diquark-triquark
configuration might also turn out to be useful if negative parity
exotic baryons are experimentally discovered in future.

\section*{Acknowledgments}

The research of one of us (M.K.) was supported in part by a grant from the
United States-Israel Binational Science Foundation (BSF), Jerusalem and
by the Einstein Center for Theoretical Physics at the Weizmann Institute.
The research of one of us (H.J.L.) was supported in part by the U.S. Department
of Energy, Division of High Energy Physics, Contract W-31-109-ENG-38,
We benefited from e-mail discussions with Ken Hicks about the experimental data on
the $\Theta^+$ and
with Simon Capstick, Frank Close, Mitya Diakonov, Bob Jaffe and Micha{\l}
Prasza{\l}owicz about the challenges this data poses for a theoretical
interpretation.
M.K. would like to thank the HEP group at Cavendish Laboratory for
hospitality.

%
\catcode`\@=11 
\def\references{
\ifpreprintsty \vskip 10ex
%
\hbox to\hsize{\hss \large \refname \hss }\else
\vskip 24pt \hrule width\hsize \relax \vskip 1.6cm \fi \list
{\@biblabel {\arabic {enumiv}}}
{\labelwidth \WidestRefLabelThusFar \labelsep 4pt \leftmargin \labelwidth
\advance \leftmargin \labelsep \ifdim \baselinestretch pt>1 pt
\parsep 4pt\relax \else \parsep 0pt\relax \fi \itemsep \parsep \usecounter
{enumiv}\let \p@enumiv \@empty \def \theenumiv {\arabic {enumiv}}}
\let \newblock \relax \sloppy
 \clubpenalty 4000\widowpenalty 4000 \sfcode `\.=1000\relax \ifpreprintsty
\else \small \fi}
\catcode`\@=12 

} 

\end{document}